\documentclass[letterpaper, 10 pt, conference]{ieeeconf}  

\IEEEoverridecommandlockouts                              

\usepackage{color}
\usepackage{graphicx} 
\usepackage{amsmath} 
\usepackage{amssymb}  
\usepackage{caption}
\usepackage{subcaption}
\usepackage{cite}

\title{\LARGE \bf
	Spacecraft Relative Motion Planning Using Chained Chance-Constrained Admissible Sets
}

\author{Andrew W. Berning Jr.$^{1}$, Nan I. Li$^{1}$, Anouck Girard$^{1}$, \\
	Frederick A. Leve$^{2}$, Christopher D. Petersen$^{3}$, and Ilya Kolmanovsky$^{1}$
	\thanks{$^{1}$Andrew W. Berning Jr., Nan I. Li, Anouck Girard, and Ilya Kolmanovsky are
		with the Department of Aerospace Engineering, The University
		of Michigan, Ann Arbor, MI 48105, USA. Emails:
		{\tt\small awbe@umich.edu}, {\tt\small nanli@umich.edu}, {\tt\small anouck@umich.edu}, and {\tt\small ilya@umich.edu}}%
	\thanks{$^{2}$ Frederick A. Leve is with The Air Force Office of Scientific Research, Arlington, VA 22203. Email: {\tt\small frederick.leve@us.af.mil}}
	\thanks{$^{3}$ Christopher Petersen is with The Air Force Research Laboratory, Albuquerque, NM 87123. Email: {\tt\small VSSVOrgMailbox@us.af.mil}}
}

\begin{document}
\maketitle
\thispagestyle{empty}
\pagestyle{empty}

\begin{abstract}
With the increasing interest in proximity and docking operations, there is a growing interest in spacecraft relative motion control. This paper extends a previously proposed constrained relative motion approach based on chained positively invariant sets to the case where the spacecraft dynamics are controlled using output feedback on noisy measurements and are subject to stochastic disturbances. It is shown that non-convex polyhedral exclusion zone constraints can be handled. The methodology consists of a virtual net of static equilibria nodes in the Clohessy-Wiltshire-Hill frame. Connectivity between nodes is determined through the use of chance-constrained admissible sets, guaranteeing that constraints are met with a specified probability. 
	
\end{abstract}

\section{Introduction}

As of April 2019, the U.S. Space Surveillance Network was tracking 19,404 pieces of orbital debris \cite{anz2019orbital}. Two major contributors to this debris are the 2007 Chinese anti-satellite missile test and the 2009 Iridium-Kosmos satellite collision, though other minor collisions contribute to the debris count yearly \cite{national2011committee}. The need to operate satellites safely in the presence of this orbital debris, as well as other operational satellites, motivates the development of relative motion planning algorithms that include nonconvex obstacle avoidance constraints. Other mission considerations include handling modeling uncertainties and measurement uncertainties while relying on the limited computational capabilities of many spacecraft. 

Spacecraft relative motion planning (SRMP) is concerned with the design of orbital maneuvers with respect to a reference point on a nominal orbit. To handle nonconvex obstacle avoidance constraints, one approach involves sequential optimization of a set of convexified problems that eventually recovers the original optimal solution \cite{liu2014solving}. Richards et al. proposes a framework in which the fuel-optimal spacecraft trajectory optimization problem subject to avoidance constraints is expressed as a mixed-integer linear program \cite{richards2002spacecraft}. A model predictive control (MPC) approach for rendezvous and proximity operation is presented in \cite{di2012model}, while \cite{baldini2016fast} approaches the SRMP problem with a computationally efficient, sampling-based algorithm. A comprehensive survey of spacecraft formation flying can be found in \cite{scharf2003survey}. 

A graph search framework for SRMP proposed in \cite{weiss2014safe, frey2017constrained, frey2018invariance} benefits from the computational efficiency and simplicity of algorithms such as Dijkstra's \cite{dijkstra1959note} and $A^*$ \cite{hart1968formal} search. The approach involves building a connectivity graph for a set of forced equilibria or natural motion trajectories and the use of safe, positively invariant sets to determine connectivity between graph vertices. The resulting motion planning framework can accommodate obstacles and bounded disturbances. 

The developments in \cite{weiss2014safe} are based on assumptions of full state measurement and set bounded disturbances.  Under these conditions, positively invariant sets are constructed around forced equilibria, which guarantee that the closed-loop response satisfies the constraints for any initial condition in this set when the selected equilibrium reference command is the one corresponding to this set. A safe transition between two forced equilibria can be accomplished if the first equilibrium is in the interior of the positively invariant set for the second equilibrium. Such forced equilibria are then treated as vertices in a directed graph (a virtual net in the terminology of \cite{weiss2014safe}) and are connected by an edge. Based on the family of such equilibria, spacecraft motion planning reduces to a graph search for the sequence of the equilibria to hop between to arrive at the target equilibrium while minimizing suitably constructed cumulative transition cost. This approach is extended in \cite{frey2017constrained} to include periodic natural motion trajectories, in \cite{frey2017incorporating} to include non-periodic trajectories, and in \cite{frey2018invariance} to handle set bounded disturbances and minimum thrust constraints. Related ideas have been explored for the development of motion planners for self-driving cars in \cite{berntorp2018positive,berntorp2017automated}.

In this paper, we consider the case when the system model is linear, the full state measurement is not available and the measured outputs are contaminated by random gaussian measurement noise.  In addition, the system is affected by random gaussian disturbances which could represent the effect of unknown forces such as thrust errors or other perturbations acting on the spacecraft. In this setting, an observer is introduced to estimate the state, and state constraints are imposed as chance (probabilistic) constraints. Note that hard constraints cannot be enforced for all times as disturbances and measurement noise values are not compactly supported. For this setting, chance constrained admissible sets are defined as sets of initial state estimates such that the chance constraints hold for all future times for the given constant reference. As the chance constraints are dependent on the initial estimation error covariance matrix, a simplifying assumption is made--the observer has reached steady-state and hence this matrix is equal to the steady-state error covariance matrix. Unlike \cite{weiss2014safe}, where positive invariant sets being chained are sublevel sets of Lyapunov functions, here we exploit chance constrained admissible sets to determine connectivity of forced equilibria. These sets are near maximal and hence allow more vertices in the graph to be connected. We also propose a novel approach of handling non-convex constraints by exploiting their inner approximation with a union of convex constraints and we extend the
connectivity conditions to this case. Further we demonstrate that the chance constraint used as a requirement in the construction of chance constrained admissible sets holds for the closed-loop trajectories with switching between forced equilibria.

This paper is organized as follows. Section \ref{sec:modeling} describes the relative motion dynamics and disturbance model. Section \ref{sec:chance_sets} describes the chance constrained admissible sets and how to construct them. Section \ref{sec:virtual_net} describes the construction of the virtual net and its extension for obstacle avoidance, while Section \ref{sec:simulations} presents some numerical examples. 

\section{Modeling} \label{sec:modeling}

Consider a system with a linear discrete-time model given by
\begin{gather}
x_{k+1} = Ax_k + Bu_k + \Gamma w_k, \label{equ:model1} \\
y_k = C x_k + F v_k, \label{equ:model2}
\end{gather}
where $x_k$ is an $n_x$-vector state, $u_k$ is an $n_u$-vector
control input, $w_k$ is an $n_w$-vector disturbance, $y_k$ is an
$n_y$-vector measured output, and $v_k$ is an $n_v$-vector measurement noise. We make the following two assumptions:

\begin{enumerate}
	\item The disturbance and noise sequences are independent and identically distributed Gaussian processes with zero mean and unit covariance matrix, i.e.,
	\begin{equation}   
	w_k \sim \mathcal{N}(0,\mathbb{I}), \quad\quad v_k \sim \mathcal{N}(0,\mathbb{I}).
	\end{equation}
	\item The variables $x_0, \{w_k\}_{k \in \mathbb{Z}_{\ge 0}}$, and $\{v_k\}_{k \in \mathbb{Z}_{\ge 0}}$ are independent.
\end{enumerate} 

The nominal dynamics model used in this work is the linearized Clohessy-Wiltshire-Hill (CWH) equations \cite{wh1960terminal}, which describe the motion of a chase spacecraft relative to a target spacecraft orbiting a central body in a circular orbit. The continuous-time CWH equations for low Earth orbit with control and parameter matrices $A$ and $B$ are as follows:

\begin{align}
A_{ct} = &~\begin{bmatrix}
0 & 0 & 0 & 1 &0&0 \\
0 & 0 & 0 & 0 &1&0 \\ 	
0 & 0 & 0 & 0 &0&1 \\ 	
3n^2 &0&0&0&2n&0 \\
0&0&0&-2n&0&0 \\
0&0&-n^2&0&0&0
\end{bmatrix}
, \\
B_{ct} =  &~\begin{bmatrix}
0_{3,3} \\
\mathbb{I}_3
\end{bmatrix}, ~
n =0.0013 \frac{\mathrm{rad}}{\mathrm{s}},
\end{align}

\noindent where $x = \begin{bmatrix}
x_1 & x_2 & x_3 & \dot{x}_1 & \dot{x}_2 & \dot{x}_3
\end{bmatrix}^\intercal$, and the value for $n$ used here corresponds to low Earth orbit. The $x_1$ axis is along the direction from the central body to the target spacecraft, the $x_3$ axis is along its angular momentum vector, and the $x_2$ axis completes the right-handed reference frame. Discretizing these equations with a zero order hold and sampling period of $\Delta T$ results in system matrices: 

\begin{align}
	A = e^{ A_{ct} \Delta T}, ~ B = \int_{0}^{\Delta T} e^{A_{ct} (\Delta T - \tau)}d \tau B_{ct}.
\end{align}

The process noise and sensor noise matrices, $\Gamma$ and $F$, and the output matrix $C$ are defined as: 

\begin{align}
\Gamma = &~ \frac{1}{100} \begin{bmatrix}
0&0&0 \\
0&0&0 \\
0&0&0 \\
1&0&0 \\
0&1&0 \\
0&0&1 
\end{bmatrix}
, ~ F = \frac{1}{100} \begin{bmatrix}
1&0&0 \\
0 &1&0 \\
0&0&1
\end{bmatrix}, \\
C = &~ \begin{bmatrix}
1 & 0 & 0 & 0 & 0 & 0 \\
0 & 1 & 0 & 0 & 0 & 0 \\
0 & 0 & 1 & 0 & 0 & 0 
\end{bmatrix} .
\end{align}

Next, a Luenberger observer of the following form is added: 
\begin{align}
\hat{x}_{k+1} &= A \hat{x}_k + Bu_k + L(\hat{y}_k-y_k), \\
& = A \hat{x}_k + Bu_k + L(C \hat{x}_k - C x_k - F v_k), \label{equ:Luenberger} 
\end{align}

\noindent as well as a feedback control law: 

\begin{align}
	u_k&=K \hat{x}_k+ G r, \label{equ:control}
\end{align}

\noindent where $r$ is the set-point. Here, $L$ and $K$ are any stabilizing gain matrices and $G$ is computed as

\begin{equation}
G = \big (C(\mathbb{I}-A-BK)^{-1}B \big )^{-1}, 
\end{equation}

\noindent so that $y=r$ in steady-state in the disturbance free case.

\section{Chance Constrained Admissible Sets} \label{sec:chance_sets}
\subsection{Covariance Computation}
Define the estimation error as
\begin{equation}\label{equ:stateerror}
e_k=x_k-\hat{x}_k.
\end{equation}

Then the estimation error dynamics are represented by the following equations,
\begin{eqnarray}\label{equ:obserror}
e_{k+1} & = & A e_k +\Gamma w_k +LC e_k +LF v_k \\ 
& = & A_o e_k+ B_o \left[\begin{array}{c} w_k \\ v_k 
\end{array}\right],    
\end{eqnarray}
where
\begin{equation}
A_o=(A+LC),\quad B_o=\left[ \begin{array}{cc}
\Gamma & LF \end{array} \right].
\end{equation}
The matrix $A_o$ is assumed to be Schur (all eigenvalues are inside the unit disk).

The evolution of the state estimate, $\hat{x}_k$, is determined from (\ref{equ:Luenberger}), (\ref{equ:control}) and (\ref{equ:stateerror}) by
\begin{equation}
\hat{x}_{k+1}=A_{c} \hat{x}_k+B_{c} r + \Gamma_c \left[\begin{array}{c} e_k \\ v_k \end{array}\right],
\end{equation}
where
\begin{align}
A_{c} =&~ (A+BK),\\
 B_{c} =&~ BG, \\
  \Gamma_c =&~ \left[\begin{array}{cc} -LC & -LF \end{array}\right].
\end{align}
The control gain $K$ is assumed to be stabilizing and the matrix $A_{c}$ is assumed to be Schur.

Let 
\begin{equation}
\tilde{x}_k=\hat{x}_k-(\mathbb{I}-A_{c})^{-1} B_{c} r. \label{equ:x_tilde}
\end{equation}
Then
\begin{gather}\label{equ:tildexdyn}
\tilde{x}_{k+1}=A_{c} \tilde{x}_k +\Gamma_c \left[\begin{array}{c} e_k \\ v_k \end{array}\right], \\[2pt]
x_k = \hat{x}_k+e_k=\tilde{x}_k + e_k + (\mathbb{I}-A_{c})^{-1}B_{c} r .
\end{gather}

The observer error dynamics (\ref{equ:obserror}) are assumed to be in steady-state and the error is assumed to be normally distributed with zero mean and steady-state covariance matrix, $P_\infty \succeq 0$, satisfying
\begin{equation} \label{eqn:P_infty}
P_\infty = A_o P_\infty A_o^\intercal+B_o B_o^\intercal.
\end{equation}
That is, $e_k \sim \mathcal{N}(0,P_\infty)$ for all $k$. This represents a situation where the closed-loop system (including the plant and the observer) has operated for a sufficiently long period of time. Note that since $A_o$ is Schur, $P_k \to P_\infty$ as $k \to \infty$ \cite{aastrom2012introduction}.

\subsection{Chance Constraints} \label{sec:chance_constraints}
Consider now enforcing a state constraint of the form
\begin{equation}\label{equ:chancecnr}
H x_k \leq h, \forall k, 
\end{equation}
where $h$ is an $n_h$-vector.  This constraint can be written using \eqref{equ:stateerror} and \eqref{equ:x_tilde} as:
\begin{equation}
H \tilde{x}_k + H e_k \leq h-H(\mathbb{I}-A_{c})^{-1}B_{c} r.
\end{equation}

We now consider approaches to enforce the constraint 
	(\ref{equ:chancecnr}) with probabilistic guarantees based on the model (\ref{equ:obserror}) and (\ref{equ:tildexdyn}).
	Re-stating the model and the constraint for convenience here, we have
	\begin{gather}
	 \left[\begin{array}{c} \tilde{x}_{k+1} \\ e_{k+1} \end{array}\right] = \underbrace{\begin{bmatrix} A_{c} & -LC \\ 0 & A_o \end{bmatrix}}_{=: A_{\text{aug}}} \left[\begin{array}{c} \tilde{x}_{k} \\ e_{k} \end{array}\right] \nonumber \\
	 + \underbrace{\begin{bmatrix} -LF \\ LF \end{bmatrix}}_{=: B_{\text{aug}}} v_{k} + \underbrace{\begin{bmatrix} 0 \\ \Gamma \end{bmatrix}}_{=: \Gamma_{\text{aug}}} w_{k}, \\[6pt]
	H \tilde{x}_k + H e_k \leq h - H(\mathbb{I}-A_{c})^{-1}B_{c} r,
	\end{gather}
	where 
	\begin{equation}
	v_k \sim \mathcal{N}(0,\mathbb{I}), \quad\quad w_k \sim \mathcal{N}(0,\mathbb{I}). 
	\end{equation}

Let $k$ be the current time instant, and consider $t \geq 0$ to be running time over the prediction horizon. Denote by $\tilde{x}_{t|k}$ the predicted value of $\tilde{x}_{k+t}$ and by $e_{t|k}$ the predicted value of $e_{k+t}$. The dynamics of $\left[\tilde{x}_{t|k},\, e_{t|k}\right]^\intercal$ are given as
	\begin{gather}
	\left[\begin{array}{c} \tilde{x}_{t+1|k} \\ e_{t+1|k} \end{array}\right] = A_{\text{aug}} \left[\begin{array}{c} \tilde{x}_{t|k} \\ e_{t|k} \end{array}\right] + B_{\text{aug}} v_{k+t} + \Gamma_{\text{aug}} w_{k+t}.
	\end{gather}
	
	We can predict the time-varying covariance matrix of $\left[\tilde{x}_{t|k},\, e_{t|k}\right]^\intercal$ using
	\begin{equation}
	\tilde{P}_{t+1|k} = A_{\text{aug}} \tilde{P}_{t|k} A_{\text{aug}}^\intercal + B_{\text{aug}} B_{\text{aug}}^\intercal + \underbrace{\Gamma_{\text{aug}} \Gamma_{\text{aug}}^\intercal}_{= \begin{bmatrix} 0 & 0 \\ 0 & \Gamma \Gamma^\intercal \end{bmatrix}},
	\end{equation}
	
	\noindent where 
	
	\begin{equation}
	\tilde{P}_{0|k} = \begin{bmatrix} 0 & 0 \\ 0 & P_\infty \end{bmatrix}.
	\end{equation}
	
	Note that $\tilde{x}_{0|k} = \tilde{x}_{k}$, as the observer output, is measured, thus, $\text{cov}(\tilde{x}_{0|k},\tilde{x}_{0|k}) = \text{cov}(\tilde{x}_{0|k},e_{0|k}) = 0$. We assume that $\text{cov}(e_{0|k},e_{0|k}) = P_{\infty}$ for all $k$, where $P_{\infty}$ is defined in \eqref{eqn:P_infty}, based on the assumption that the closed-loop system including the observer has operated for a sufficiently long period of time.
	
	Then,
	\begin{equation}
	\left[\begin{array}{c} \tilde{x}_{t|k} \\ e_{t|k} \end{array}\right] \sim \mathcal{N}\bigg(A_{\text{aug}}^t \left[\begin{array}{c} \tilde{x}_{k} \\ 0 \end{array}\right],\tilde{P}_{t|k}\bigg),
	\end{equation}
	so
	\begin{align}
	\underbrace{\big[\, H\,\, H\, \big]}_{=:H_{\text{aug}}} \left[\begin{array}{c} \tilde{x}_{t|k} \\ e_{t|k} \end{array}\right] \sim \mathcal{N}\bigg(H_{\text{aug}} A_{\text{aug}}^t \left[\begin{array}{c} \tilde{x}_{k} \\ 0 \end{array}\right], \nonumber \\
	 \underbrace{H_{\text{aug}}\tilde{P}_{t|k}H_{\text{aug}}^\intercal}_{=:\Sigma_{t|k}}\bigg).
	\end{align}
	
	Due to the fact that the disturbance and noise signals are unbounded, it is in general not possible to enforce the constraint \eqref{equ:chancecnr} for all possible realizations of disturbance and noise sequences. Therefore, we instead consider a chance constraint imposed over the prediction horizon of the form
	\begin{equation}\label{equ:originalchanceconstraint}
	\text{Prob} \{H x_{t|k} \leq h\} \geq 1-\alpha,\quad t \geq 0,
	\end{equation}
	where $0 < \alpha < 1$.
	This chance constraint can be re-stated as
	\begin{align}\label{equ:probcnr}
\text{Prob} \bigg\{ H_{\text{aug}}\left[\begin{array}{c} \tilde{x}_{t|k} \\ e_{t|k} \end{array}\right] \leq h - H(\mathbb{I}-A_{c})^{-1} B_{c} r &\bigg\}  \\
\geq 1-\alpha&,\quad t \geq 0. \nonumber
	\end{align}

Assume for the moment that the constraint is scalar, $n_h=1$ (this assumption will be relaxed using a risk allocation approach later in this section). In this case, the constraint (\ref{equ:probcnr}) can be re-stated as
\begin{align}
\label{equ:detcnr}
H_{\text{aug}} A_{\text{aug}}^t \left[\begin{array}{c} \tilde{x}_{k} \\ 0 \end{array}\right] \leq h - H(\mathbb{I}-A_{c})^{-1} B_{c} r \nonumber \\
- \sqrt{2\Sigma_{t|k}}\, {\rm erf}^{-1}(1-2\alpha),
\end{align} 
where
\begin{equation}
{\rm erf}(x)=\frac{2}{\sqrt{\pi}} \int_0^x e^{-t^2} dt.
\end{equation}

Motivated by the above considerations, define the chance constrained admissible set, $\tilde{O}_\infty(r)$, as
\begin{align}
\tilde{O}_\infty(r) = \bigg\{\tilde{x}_0:H_{\text{aug}} A_{\text{aug}}^t \left[\begin{array}{c} \tilde{x}_{0} \\ 0 \end{array}\right] \leq h(r) \nonumber \\
- \sqrt{2\Sigma_{t|0}}\, {\rm erf}^{-1}(1-2\alpha),\, t \geq 0 \bigg\}, \label{eqn:O_infty}
\end{align}
where 
\begin{equation}
h(r) = h - H(\mathbb{I}-A_{c})^{-1} B_{c} r. \label{eqn:h_r}
\end{equation}

For numerical implementation, $\tilde{O}_\infty(r)$ is constructed as described in \cite{kalabic2019reference}. For $n_h \ge 1$, let $H_i$ denote the $i$th row of $H$ and $h_i$ denote the $i$th entry of $h$. Based on Boole's inequality, the chance constraint \eqref{equ:originalchanceconstraint} can be satisfied by enforcing the following set of constraints:
\begin{equation}
\text{Prob} \{H_i x_{t|k} \le h_i \} \ge 1- \alpha',
\end{equation}
for all $i = 1,...,n_h$, where $\alpha' = \frac{\alpha}{n_h}$. Then, we define $\tilde{O}_{\infty}(r)$ as
\begin{equation}
\tilde{O}_{\infty}(r) = \bigcap_{i=1}^{n_h} \tilde{O}_{\infty,i}(r),
\end{equation}
where $\tilde{O}_{\infty,i}(r)$ is defined using \eqref{eqn:O_infty} with $H$, $h$, and $\alpha$ replaced by, respectively, $H_i$, $h_i$, and $\alpha'$.

\section{Virtual Net} \label{sec:virtual_net}

\subsection{Graph Construction}

In its simplest form, the virtual net constrained motion planning framework exploits a discrete set of set-points,
$\mathcal{R}=\{r^1,~r^2,\cdots,r^{n_r}\},$ and reduces the trajectory design problem to an online graph search for the path in this set of set-points.  Once the path is determined through the graph search, an online switching logic is used to execute the path whereby a switch from one set-point to the next is effected when suitable switching conditions are satisfied.

The virtual net guarantees safety (constraint enforcement) by declaring that a connection (edge) between set-points $r^i$ and $r^j$ exists if  
\begin{equation}\label{equ:connectivitycon}
(\mathbb{I}-A_{c})^{-1} B_{c} (r^i - r^j) \in int\big(\tilde{O}_\infty(r^j)\big),
\end{equation} 

\noindent i.e., a connection between set-points $r^i$ and $r^j$ exists if the vector from the disturbance-free equilibrium of $r^j$ to the disturbance-free equilibrium of $r^i$ lies within the chance-constrained admissible set of $r^j$. This ensures that there will necessarily be some $k$ for which a safe switch between set-points $r^i$ and $r^j$ is possible. Specifically, suppose the system has been operating with the set-point $r^i$ for a while.  Then the dynamics of $\tilde{x}_k$ have been evolving according to (\ref{equ:tildexdyn}) with $r=r^i$. Under the assumptions made, $\hat{x}_k$ will enter an arbitrary small neighborhood of the disturbance-free equilibrium, 
$(\mathbb{I}-A_{c})^{-1} B_{c} r^i$, for some $k$. In particular, the condition,
\begin{equation}
\hat{x}_k - (\mathbb{I}-A_{c})^{-1} B_{c} r^j \in \tilde{O}_\infty(r^j)
\end{equation}
is guaranteed to hold for some $k$.  If this condition holds, then the switch of the set-point $r^i$ to $r^j$ can be effected at the time instant $k$ while guaranteeing that if the set-point $r_k$ is maintained at $r^j$ for all the subsequent time instants, the constraint
(\ref{equ:detcnr}) and hence the chance constraint (\ref{equ:originalchanceconstraint}) will be satisfied. These properties of the framework will be formally presented as Propositions~1 and~2.

The online switching controller monitors the state estimate, $\hat{x}_k$, and checks whether for the next set-point in the path, $r^+$, the switching condition,
\begin{equation}\label{equ:switchingcon}
\hat{x}_k - (\mathbb{I}-A_{c})^{-1} B_{c} r^+ \in \tilde{O}_\infty(r^+),
\end{equation} 
is satisfied.  Once (\ref{equ:switchingcon}) holds, the switch $r_k \leftarrow r^+$ is made.

With the goal of generating fuel-efficient trajectories, the graph weighting from node $r^i$ to node $r^j$, $\mathcal{G}(i,j)$ is defined as the approximate fuel needed for the spacecraft to travel from node $r^i$ to node $r^j$. The model presented in \eqref{equ:model1} with control \eqref{equ:control} is propagated subject to zero disturbance and perfect observations, i.e., $\Gamma = 0$ and $\hat{x} = x$. The initial condition is set as $x_0 = \begin{bmatrix}
r^i \\
0
\end{bmatrix}$ and the reference is set as $r = r^j$. The system is then propagated until 

\begin{align}
	\left\lvert (\mathbb{I}-A_{c})^{-1} B_{c} r^j - x_k \right\rvert_2 \leq  \nonumber \\
	0.05 \left\lvert (\mathbb{I}-A_{c})^{-1} B_{c} (r^j - r^i) \right\rvert_2
\end{align}

\noindent and the graph weighting is set as
\begin{equation}
\mathcal{G}(i,j) = \sum_{i=0}^{k} \left\lvert u_i \right\rvert_2.  
\end{equation}

{\bf Proposition~1:} Suppose that the initial pair of state estimate and set-point $(\hat{x}_0,r_0)$ satisfies $\hat{x}_0 - (\mathbb{I}-A_c)^{-1}B_c r_0 \in \tilde{O}_{\infty}(r_0)$, and $w_k,v_k \sim \mathcal{N}(0,\mathbb{I})$, $e_{0|k} \sim \mathcal{N}(0,P_\infty)$ for all $k \ge 0$. And suppose that all of the set-point switches are made when the switching condition (45) is satisfied. Then, the probability of satisfying the constraint (24) is higher than $1-\alpha$, i.e., $\text{Prob}(Hx_k \le h) \ge 1- \alpha$, for all $k \ge 0$.

{\bf Proof:} For any $k \ge 0$, let $k' = \max \big(\{t \,|\, 1 \le t \le k, r_t \neq r_{t-1}\} \cup \{0\}\big)$. Note that $k'$ is a random variable. According to the definition of $k'$ and the set-point switching condition (45), for any realization of $k'$, the corresponding set of realizations of state estimate and set-point trajectory $\{(\hat{x}_t,r_t)\}_{t=0}^{k}$ must all satisfy $\hat{x}_{k'} - (\mathbb{I}-A_c)^{-1}B_c r_{k'} \in \tilde{O}_{\infty}(r_{k'})$ and $r_t = r_{k'}$ for all $t = k',\dots,k$. Then, by the definition of $\tilde{O}_{\infty}(r_{k'})$, the conditional probability measure of the subset of trajectories satisfying $Hx_k \le h$ must be greater than $1- \alpha$, i.e., $\text{Prob}(Hx_k \le h \,|\, k' ) \ge 1- \alpha$, where $\text{Prob}(\cdot \,|\, k' )$ denotes the probability measure conditioned on the realized $k'$. Then, using the formula of total probability, we obtain $\text{Prob}(Hx_k \le h) = \sum_{k'=0}^{k} \text{Prob}(Hx_k \le h \,|\, k')\text{Prob}(k') \ge (1- \alpha) \sum_{k'=0}^{k}\text{Prob}(k') = 1- \alpha$, since $\sum_{k'=0}^{k}\text{Prob}(k') = 1$. $\blacksquare$

{\bf Proposition~2:} For a path determined by the graph search algorithm, as a sequence of set-points $\{r^0,r^1,...,r^f\}$ satisfying
$$
(\mathbb{I}-A_{c})^{-1} B_{c} (r^{i-1}-r^{i}) \in int\big(\tilde{O}_{\infty}(r^i)\big)
$$
for all $i=1,...,f$, suppose that $r_0 = r^0$ and set-point switches $r_k \leftarrow r^+$ are made when the switching condition \eqref{equ:switchingcon} is satisfied. Then, there almost surely exists $k^f \in \mathbb{N}$ such that $r_k = r^f$ for all $k \ge k^f$, i.e., the terminal set-point of the path $r^f$, as the reference point for the spacecraft to track, is reached by $r_k$ in finite time. 

{\bf Proof:} For any $i = 1,...,f$, since $(\mathbb{I}-A_{c})^{-1} B_{c} (r^{i-1}-r^{i}) \in int\big(\tilde{O}_{\infty}(r^i)\big)$, there exists an open set $U$ containing $0$ such that $U + (\mathbb{I}-A_{c})^{-1} B_{c} r^{i-1} \subset \tilde{O}_{\infty}(r^i) + (\mathbb{I}-A_{c})^{-1} B_{c} r^{i}$. Considering the system \eqref{equ:tildexdyn}, by the fact that $A_{c}$ is strictly Schur, for any initial condition $\tilde{x}_0$, there almost surely exists $k' \in \mathbb{N}$ such that $\tilde{x}_{k'} \in U$ \cite{kalabic2019reference}. This implies that if $r_k = r^{i-1}$ for a sufficiently long period of time, there almost surely exists $k' \in \mathbb{N}$ such that $\hat{x}_{k'} \in U + (\mathbb{I}-A_{c})^{-1} B_{c} r^{i-1} \subset \tilde{O}_{\infty}(r^i) + (\mathbb{I}-A_{c})^{-1} B_{c} r^{i}$, where the set-point switching condition \eqref{equ:switchingcon} is satisfied and thus $r_{k'} \leftarrow r^i$. Then, the statement of Proposition~2 follows from the fact that the above result holds for all $i = 1,...,f$. $\blacksquare$

\subsection{Obstacle Avoidance }

We have a framework in which the spacecraft is guaranteed to satisfy the chance constraints of the form \eqref{equ:originalchanceconstraint}, but the extension to obstacle avoidance necessitates non-convex keep-out zones that cannot be expressed in the form $H x_k \leq h$. In this work, we consider a scenario in which the spacecraft's motion is constrained to be inside a set defined by $H x_k \leq h$ and outside of the obstacle defined by $Q x_k \leq q$. 

\begin{figure}
	\centering
	\begin{subfigure}{.32\linewidth}
		\centering
		\includegraphics[width=\linewidth]{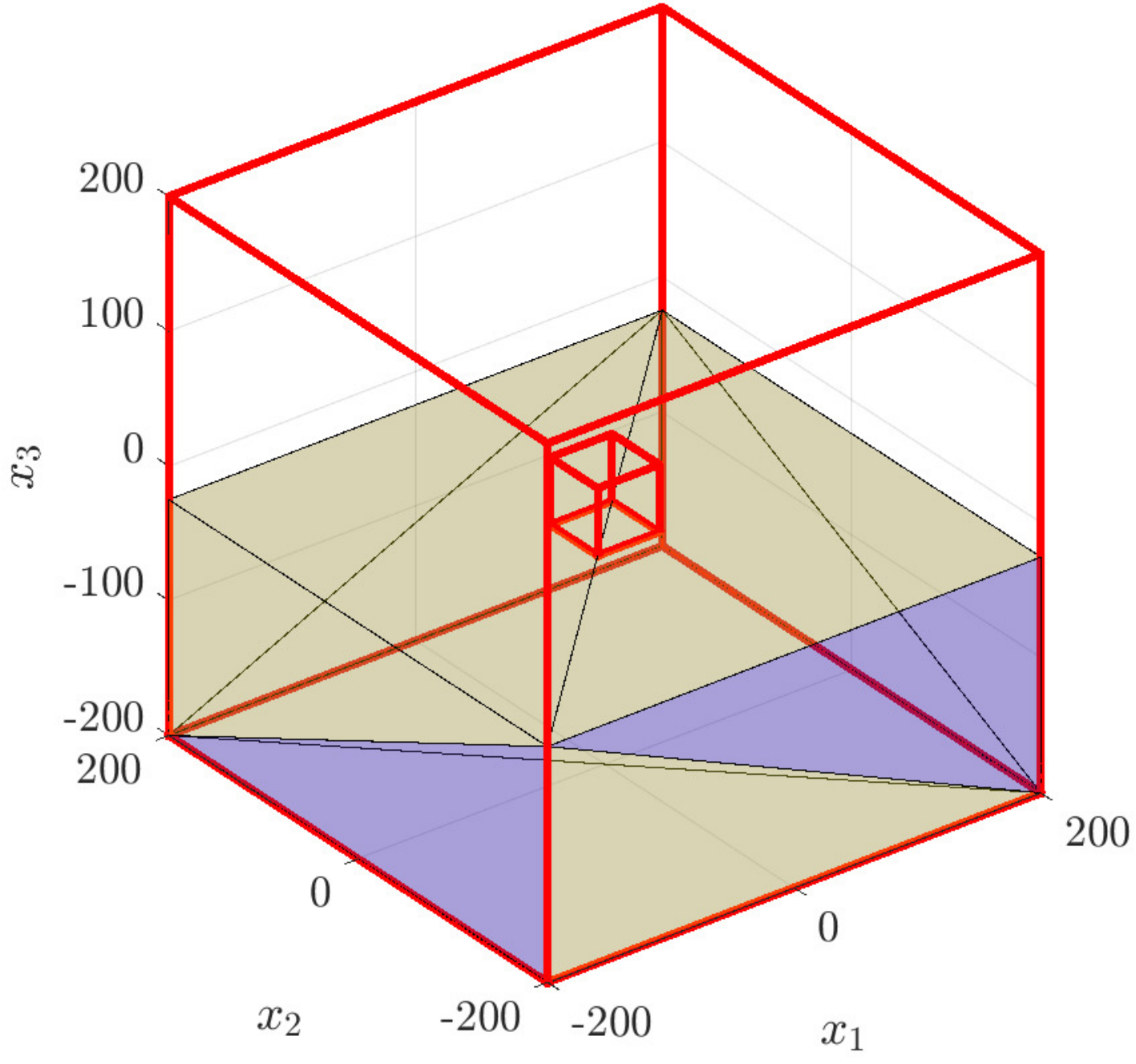}
	\end{subfigure} \hfill
	\begin{subfigure}{.32\linewidth}
		\centering
		\includegraphics[width=\linewidth]{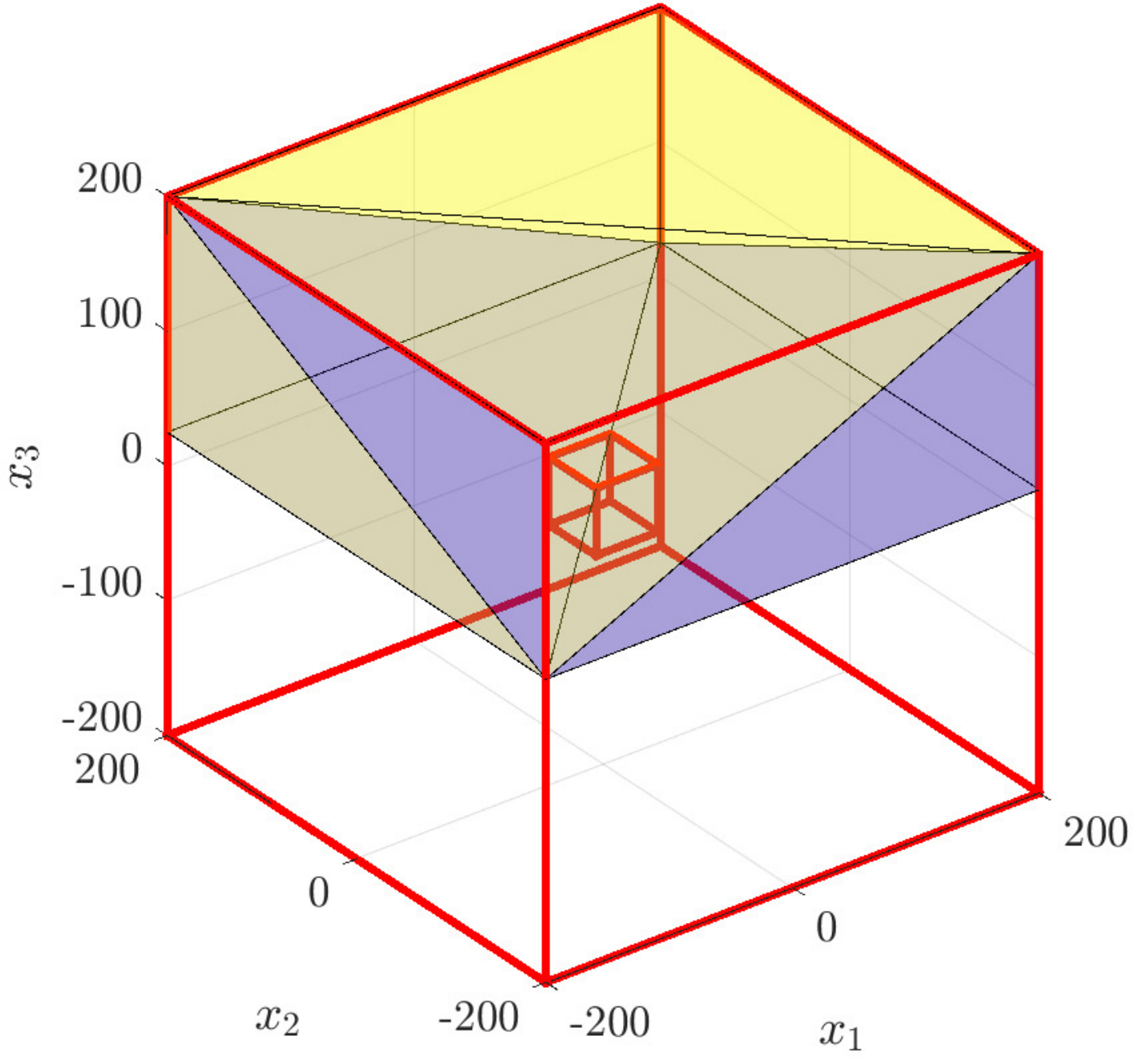}
	\end{subfigure} \hfill
	\begin{subfigure}{.32\linewidth}
		\centering
		\includegraphics[width=\linewidth]{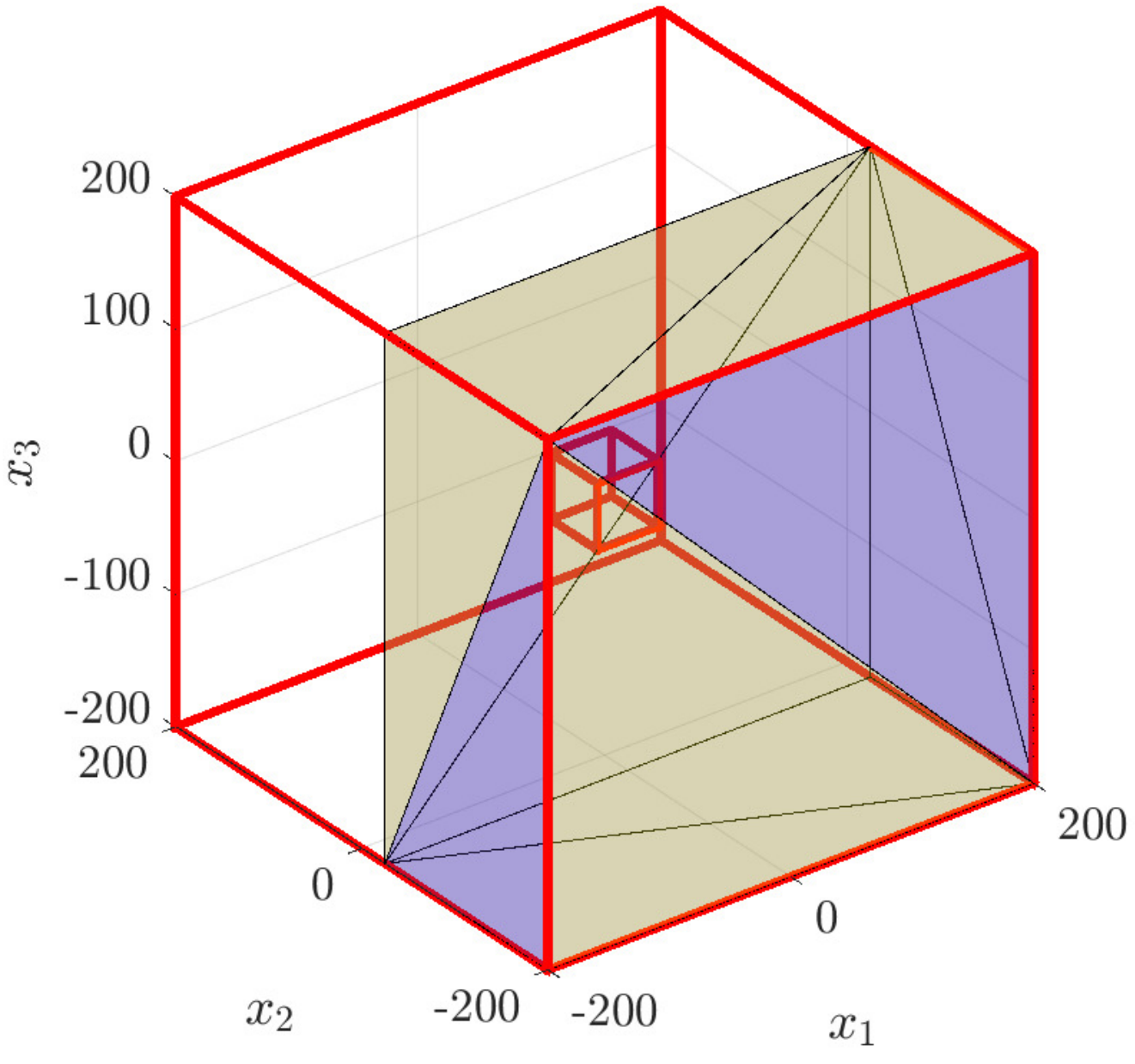}
	\end{subfigure} \hfill

	\centering
	\begin{subfigure}{.32\linewidth}
		\centering
		\includegraphics[width=\linewidth]{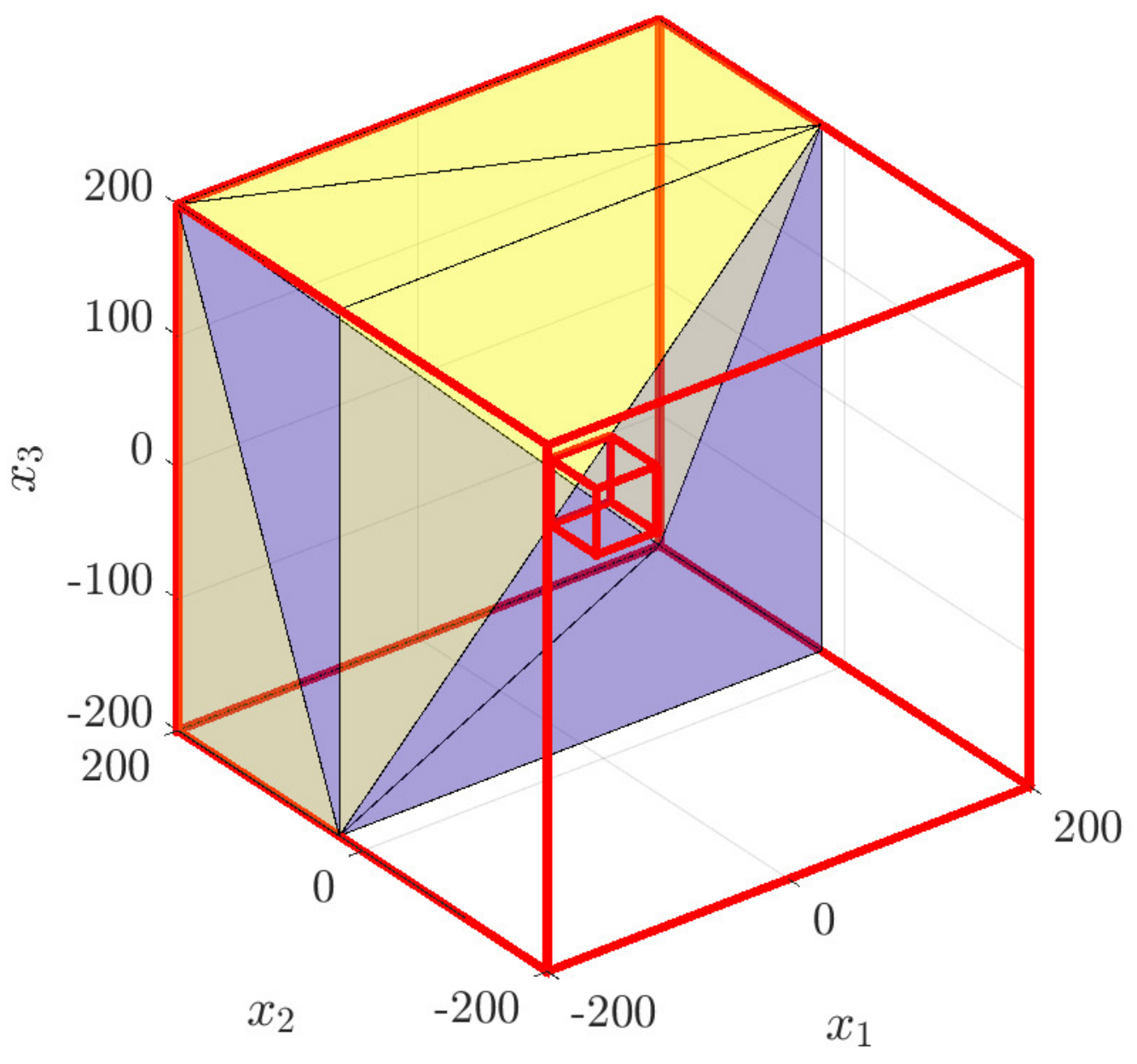}
	\end{subfigure} \hfill
	\begin{subfigure}{.32\linewidth}
		\centering
		\includegraphics[width=\linewidth]{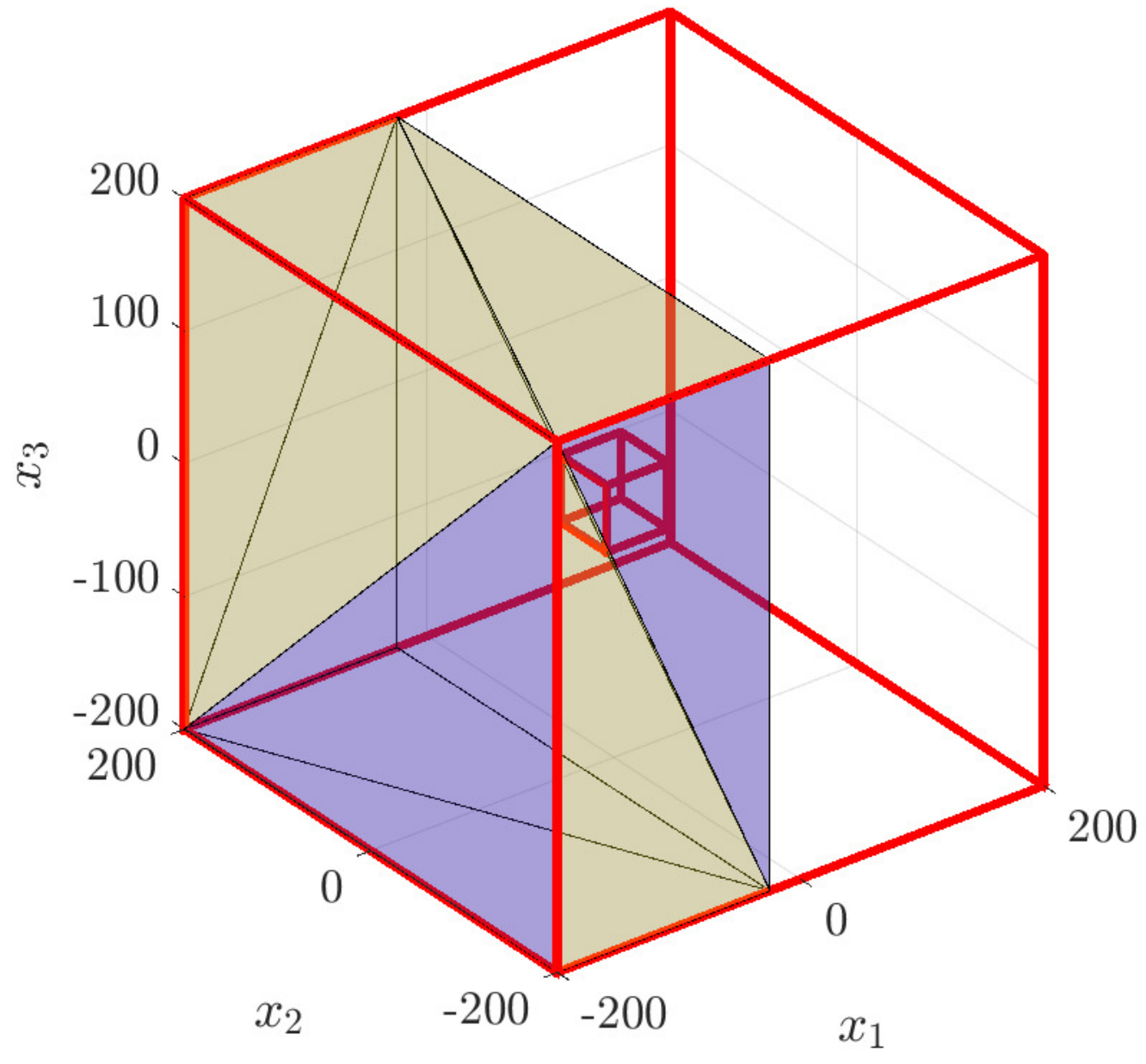}
	\end{subfigure} \hfill
	\begin{subfigure}{.32\linewidth}
		\centering
		\includegraphics[width=\linewidth]{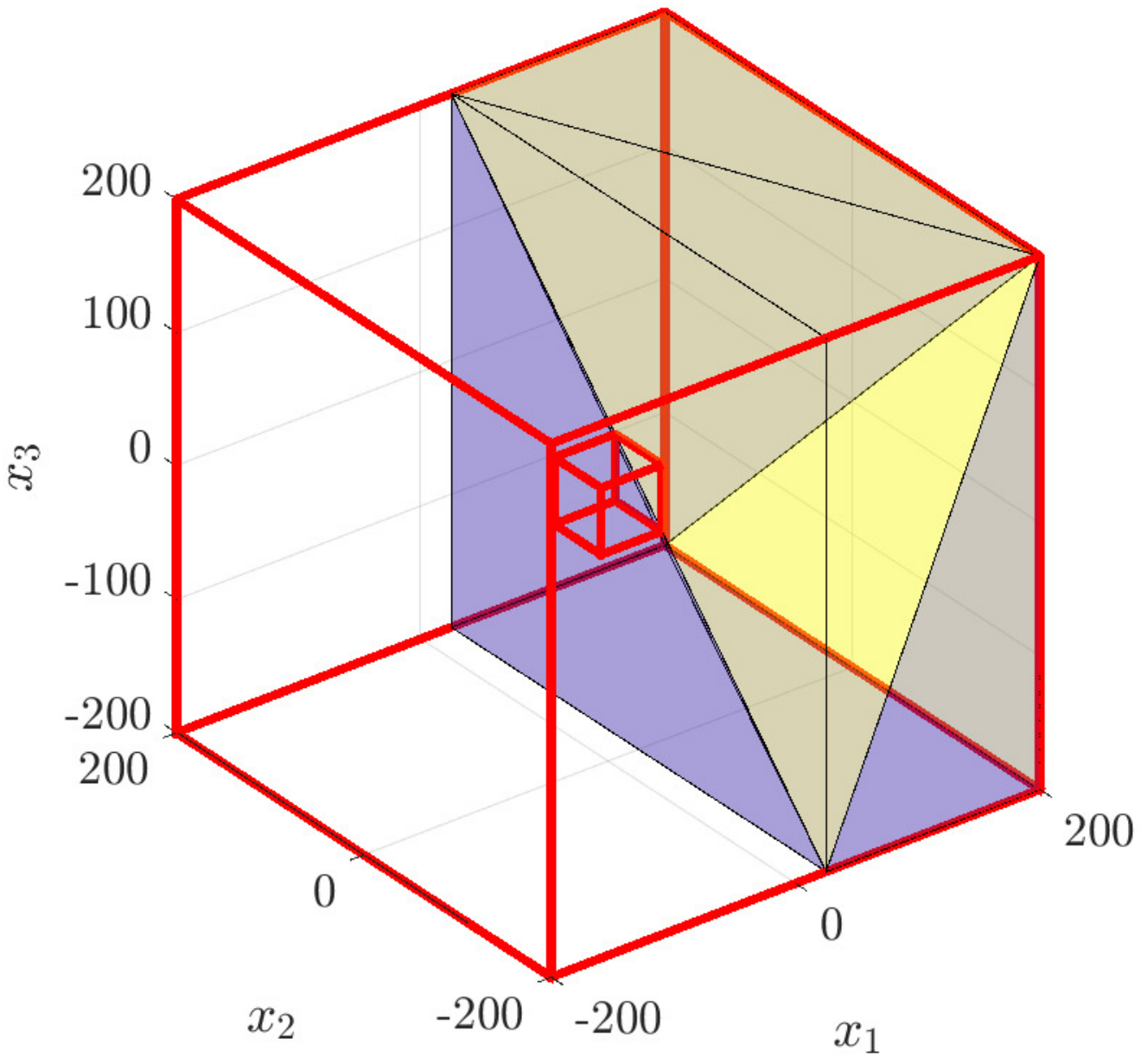}
	\end{subfigure} \hfill
	\caption{Inner approximation of a non-convex set $\mathbb{C}_{nc}$ by a union $\mathbb{C}_{c}$ of convex sets.}
	\label{fig:nonconvex2convex}
\end{figure}

This problem is solved by an inner approximation of the non-convex set $\mathbb{C}_{nc} = \{ ~ \xi \mid   H \xi \leq h  ~\} \setminus \{ ~ \xi \mid   Q \xi \leq q  ~\}$ by a union of $N_s$ convex sets $\mathbb{C}_{c} = \bigcup\limits_{i=1}^{N_s} \{ ~ \xi \mid   H_i \xi \leq h_i  ~\}$ such that $\mathbb{C}_{c} \subseteq \mathbb{C}_{nc}$. This is illustrated in Figure \ref{fig:nonconvex2convex}, which depicts an example with a cube obstacle inside outer box constraints. For each of these new sets $\mathbb{C}_{c,i}= \{ ~ \xi \mid H_i \xi \leq h_i  ~\}, i=1,\dots,N_s$, a chance-constrained admissible set may be defined as in \eqref{eqn:O_infty} and \eqref{eqn:h_r},

\begin{align}
\tilde{O}_{\infty,i}(r) = \bigg\{\tilde{x}_0:\big[\, H_i\,\, H_i\, \big] A_{\text{aug}}^t \left[\begin{array}{c} \tilde{x}_{0} \\ 0 \end{array}\right] \leq h_i(r) \nonumber \\
- \sqrt{2\Sigma_{t|0}}\, {\rm erf}^{-1}(1-2\alpha),\, t \geq 0 \bigg\}, \label{eqn:O_infty_i}
\end{align}
where 
\begin{equation}
h_i(r) = h_i - H_i(\mathbb{I}-A_{c})^{-1} B_{c} r. \label{eqn:h_r_i}
\end{equation}

Thus a connection between set-points $r^i$ and $r^j$ exists if
\begin{equation}\label{equ:connectivitycon_i}
(\mathbb{I}-A_{c})^{-1} B_{c} (r^i - r^j) \in int\big(\tilde{O}_{\infty,i}(r^j)\big).
\end{equation} 
 \noindent for any $i = 1, \dots, N_s$. 

{\bf Corollary 1:} Suppose that $\hat{x}_0 - (\mathbb{I}-A_{c})^{-1}B_{c} r_{0} \in \tilde{O}_{\infty,i}(r_{0})$ for some $i = 1,...,N_s$, and $w_k,v_k \sim \mathcal{N}(0,\mathbb{I})$, $e_{0|k} \sim \mathcal{N}(0,P_\infty)$ for all $k \ge 0$. And suppose that all of the set-point switches $r_k \leftarrow r^+$ are made when the switching condition $\hat{x}_k - (\mathbb{I}-A_{c})^{-1}B_{c} r^+ \in \tilde{O}_{\infty,i}(r^+)$ is satisfied for some $i = 1,...,N_s$. Then, the probability of staying in the safety set $\mathbb{C}_{nc}$ is higher than $1-\alpha$, i.e., $\text{Prob}(x_k \in \mathbb{C}_{nc}) \ge 1- \alpha$, for all $k \ge 0$.

{\bf Proof:} By a similar proof as that for Proposition~1, it can be shown $\text{Prob}(x_k \in \mathbb{C}_{c}) \ge 1- \alpha$ for all $k \ge 0$. Then, the statement $\text{Prob}(x_k \in \mathbb{C}_{nc}) \ge 1- \alpha$ for all $k \ge 0$ follows from the fact that $\mathbb{C}_{c} \subseteq \mathbb{C}_{nc}$. $\blacksquare$

\section{Simulations} \label{sec:simulations}

The simulation case studies presented here use the model described in Section \ref{sec:modeling}. Gain matrices $K$ and $L$ are computed by solving the discrete-time algebraic Riccati equation with corresponding weighting matrices $Q_K=10 ^{-7} \mathbb{I}_6$, $R_K=10 \mathbb{I}_3$, $Q_L=10 ^{-7} \mathbb{I}_6$, and $R_L=\mathbb{I}_3$. The discrete-time sampling period and chance constraint probability used for simulation are $\Delta T = 10 s$ and $\beta = 1 - \alpha = 0.9 $, respectively. The obstacle used for this study is a 9-sided pyramid emanating from the origin, which is intended to be an analogue for a sensor-based keep-out zone. In this scenario, it is envisioned that our chaser spacecraft has an optical sensor that is constantly pointing at the target spacecraft at the origin, and the cone represents the region in which the sensor would be damaged by the Sun. 

\begin{figure}[htbp!]
	\centering
	\includegraphics[width=\linewidth]{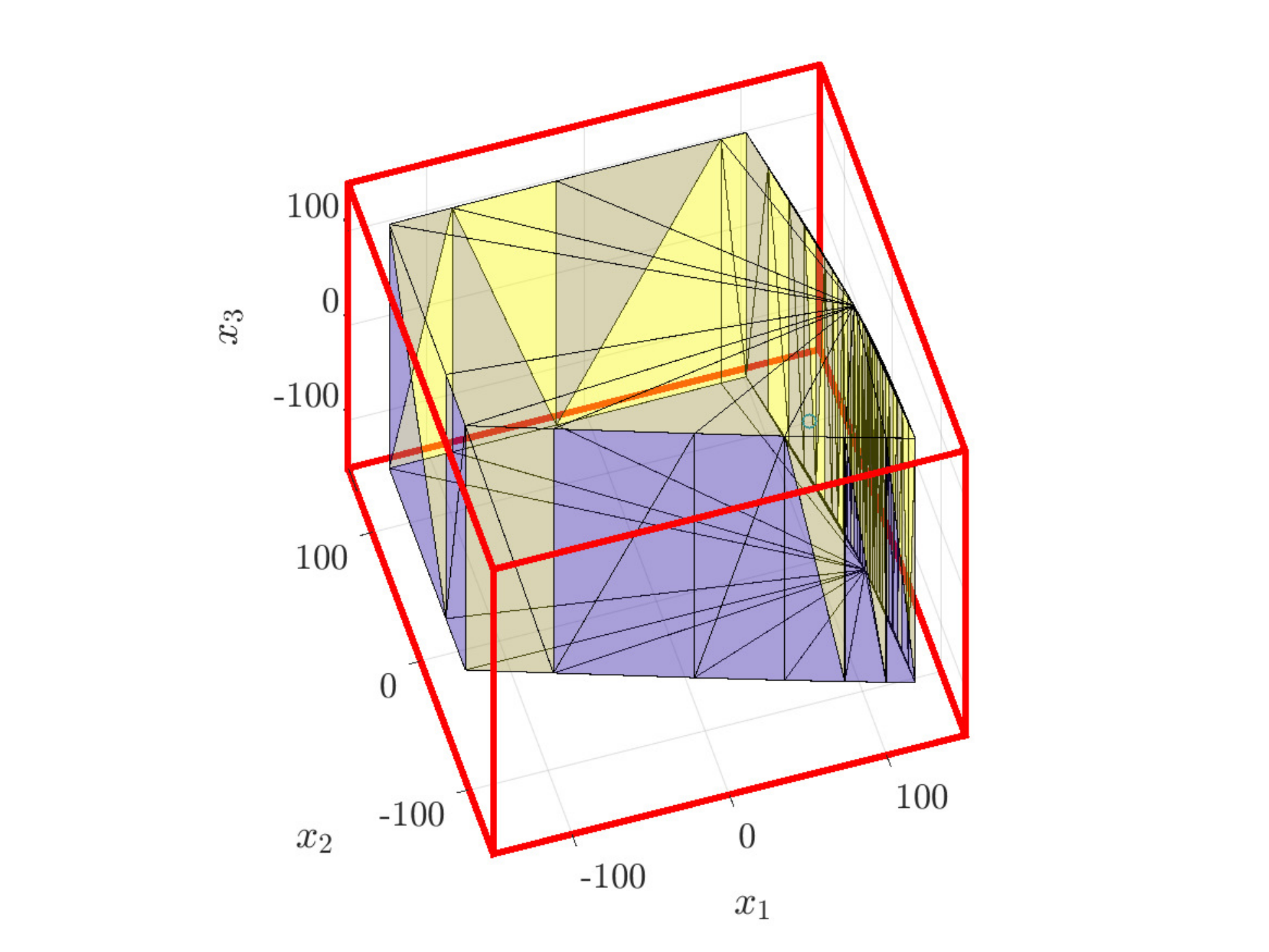}
	\caption{Visualization of $\tilde{O}_\infty([97,~ 0,~ 0]^\top)$ for the cuboid constraints shown in red.}
	\label{fig:o_inf}
\end{figure}

\begin{figure}
	\begin{subfigure}{.47\linewidth}
		\centering
		\includegraphics[width=\linewidth]{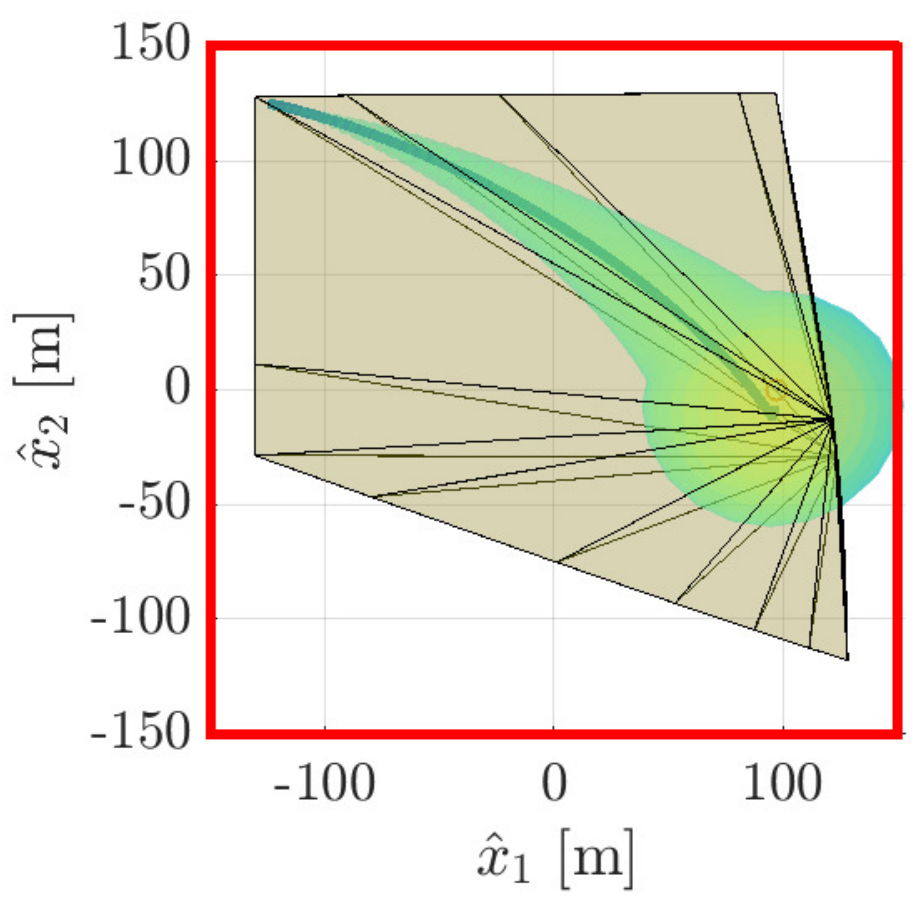}
		\caption{Example trajectory where $x_0 \in \tilde{O}_\infty$ and the $\beta$ chance constraints are satisfied, as illustrated by the $\beta$-probability trajectory tube residing fully within the box constraint. }
	\end{subfigure} \hfill
	\begin{subfigure}{.47\linewidth}
		\centering
		\includegraphics[width=\linewidth]{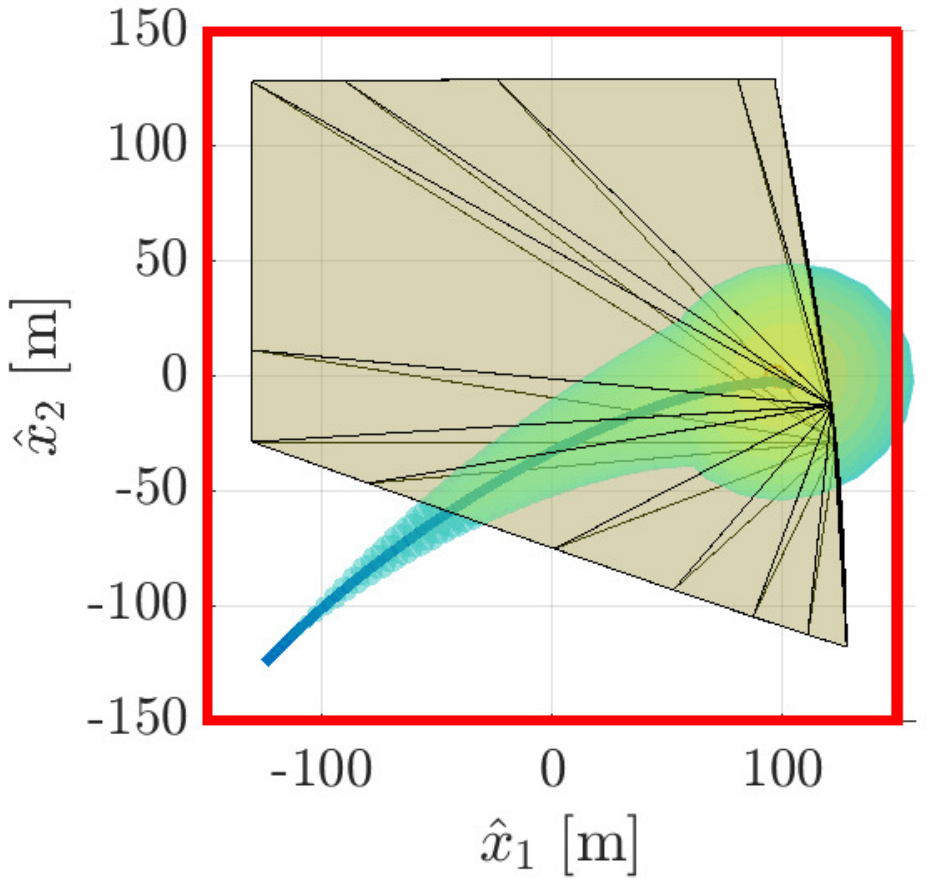}
		\caption{Example trajectory where $x_0 \notin \tilde{O}_\infty$ and the $\beta$ chance constraints are \textit{not} satisfied, as illustrated by the $\beta$-probability trajectory tube extending past the box constraint.}
	\end{subfigure}
	\caption{Comparison of safe and non-safe trajectories, as shown by the $\tilde{O}_\infty$ set.}
	\label{fig:safe_traj_comp}
\end{figure}

Figure \ref{fig:o_inf} depicts an example of one particular chance-constrained admissible set, $\tilde{O}_\infty([97,~ 0,~ 0]^\top)$. Note that this set does not extend all the way to box constraints and that it is non-symmetric about the $x_1$ and $x_2$ axes. Figure \ref{fig:safe_traj_comp} gives an intuitive explanation of these features. The same $\tilde{O}_\infty$  set is shown projected onto the $x_1{-}x_2$ plane and the trajectory tubes from two separate simulations are overlaid: one where $x_0 \in \tilde{O}_\infty$ and one where $x_0 \notin \tilde{O}_\infty$. The former shows that the box chance constraints are satisfied while in the latter simulation they are not, illustrating how the asymmetries in the relative orbital dynamics manifest in the asymmetric $\tilde{O}_\infty$ set.

\begin{figure}[htbp!]
	\centering
	\includegraphics[width=\linewidth]{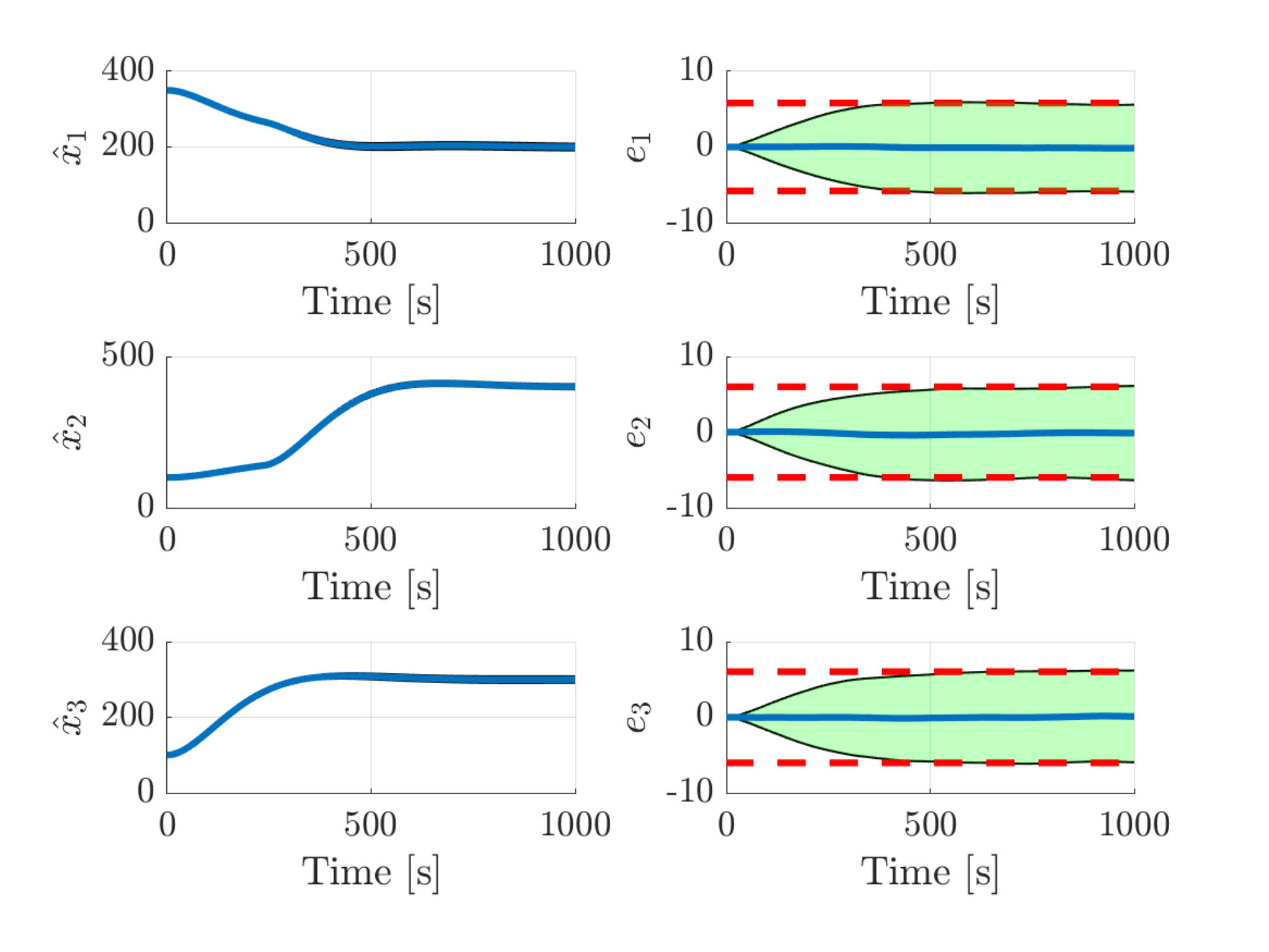}
	\caption{Position estimated states $\hat{x}$ and position error states $e$ from Monte Carlo simulations. Green shaded region is the computed $\beta$-probability distribution and the dashed red line is the $\beta$-probability distribution predicted by \eqref{eqn:P_infty}.}
	\label{fig:position_states}
\end{figure}

\begin{figure}
	\begin{subfigure}{.47\linewidth}
		\centering
		\includegraphics[width=\linewidth]{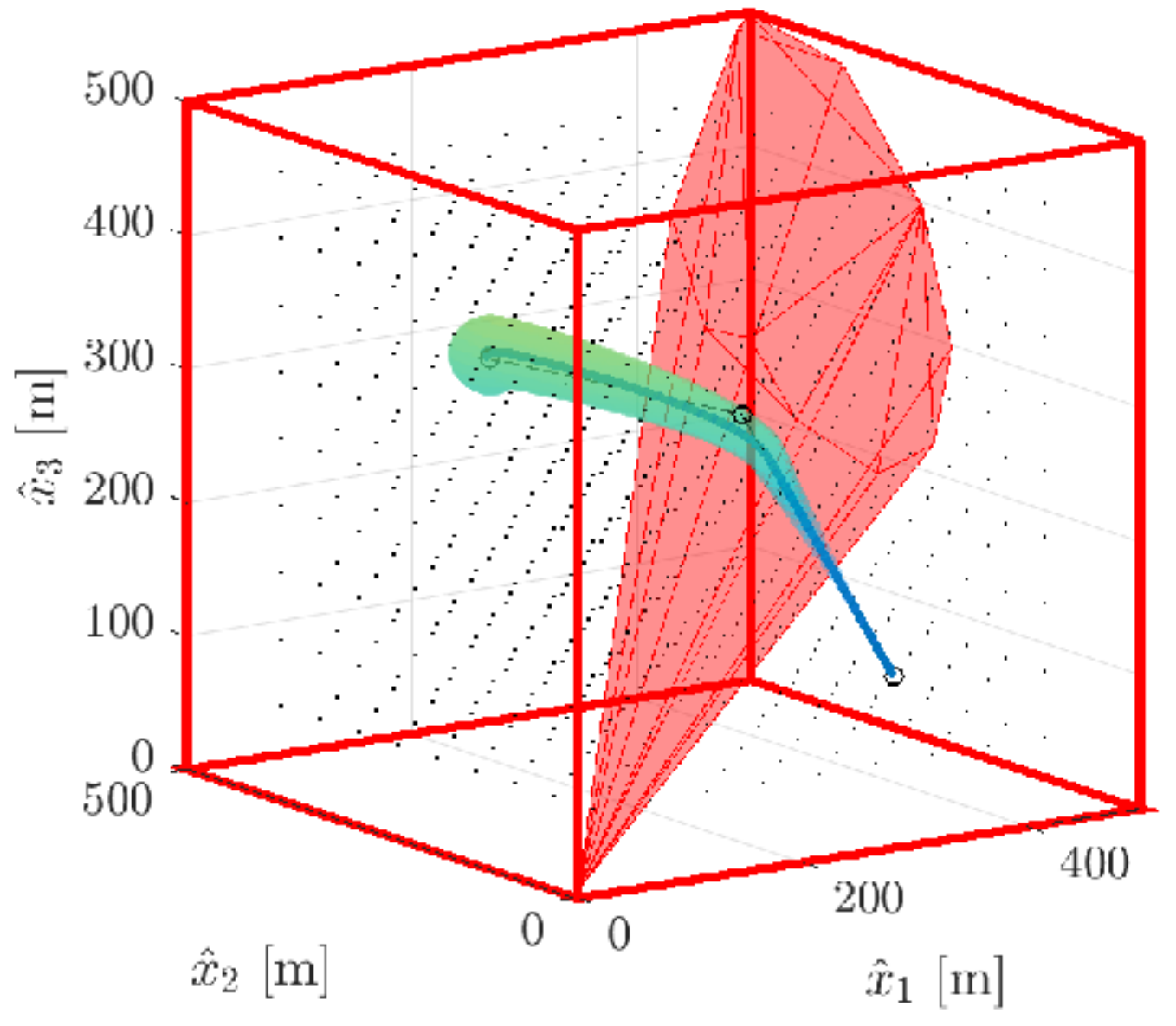}
		\caption{View 1}
	\end{subfigure} \hfill
	\begin{subfigure}{.47\linewidth}
		\centering
		\includegraphics[width=\linewidth]{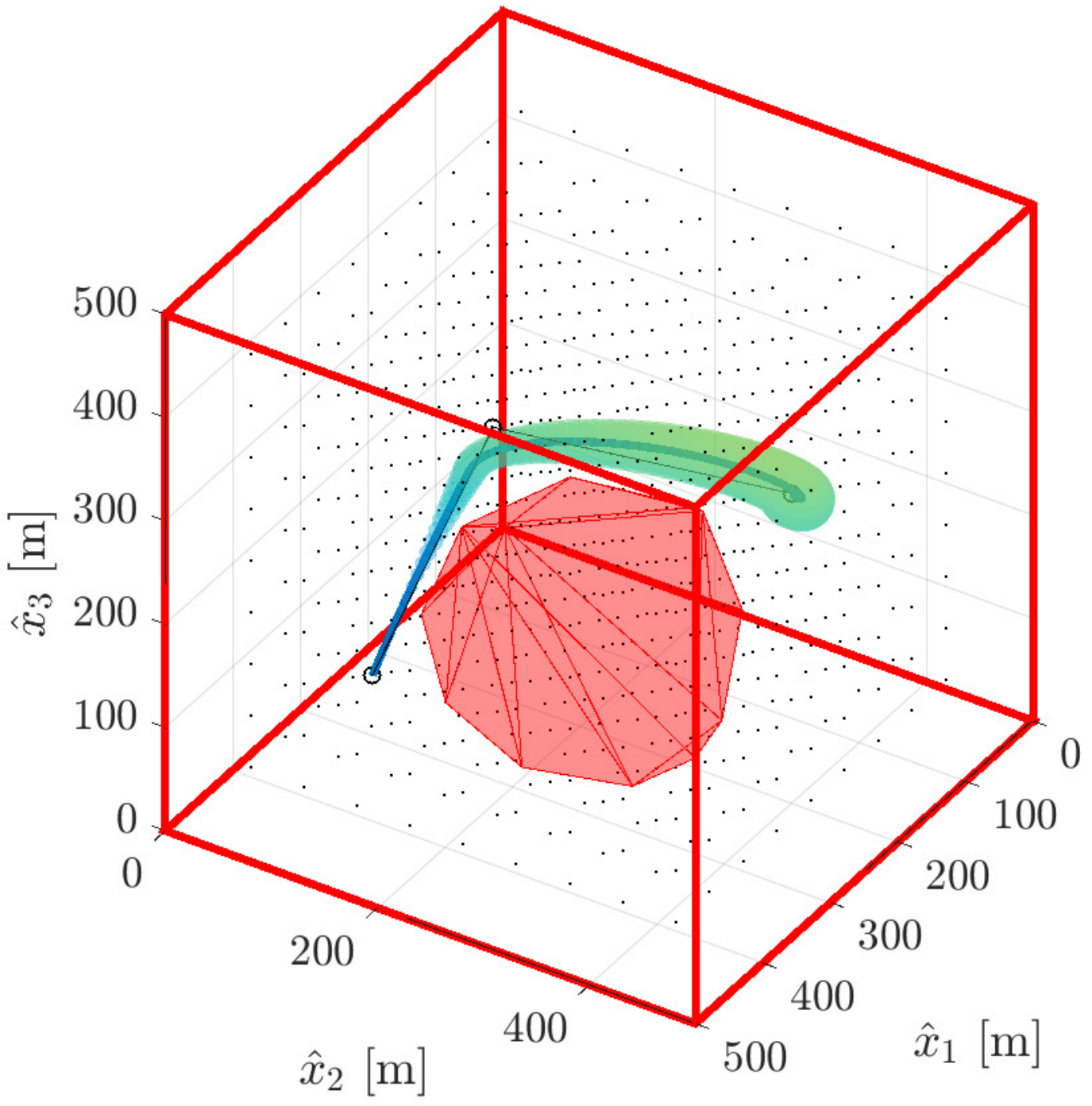}
		\caption{View 2}
	\end{subfigure}
	\caption{Three-dimensional trajectory tracking $\mathcal{R}$ (circular markers), while avoiding the red conical obstacle. The dot markers are the nodes of the virtual net.}
	\label{fig:3dtrajs}
\end{figure}

After the graph $\mathcal{G}$ is constructed, the standard Dijkstra's algorithm is used for path planning and the solution trajectory is propagated for a 1,000 run Monte Carlo simulation. Figures \ref{fig:position_states} and \ref{fig:3dtrajs} show the resulting $\beta$-probability trajectory tube, which is defined as the union of $\beta$-probability ellipsoids 

\begin{equation} \label{eq:prob_ellipse}
\{ ~ \omega \in \mathbb{R} \mid (\omega - x_{1:3,k})^{\intercal} P_{1:3,k}^{-1}   (\omega - x_{1:3,k})\leq c^2 ~\} ,
\end{equation}  

\noindent where $c$ is solved for using the three degree-of-freedom chi-squared distribution \cite{lancaster1969chi}.

In these simulations, the spacecraft successfully navigates from $x_0$ to the final reference point $r_f$ while satisfying the chance state constraints and avoiding the obstacle. Additionally, it is shown that the experimentally computed covariance of $e$ matches the theoretical value of $P_\infty$ from \eqref{eqn:P_infty}.

\section{Conclusions} \label{sec:conclusions}

In this paper, the relative motion planning framework based on chained positively invariant constraint admissible sets in \cite{weiss2014safe} was extended to the setting of stochastic disturbances and output measurement with stochastic measurement noise. With the proposed approach, chance constraints are considered and maximal chance-constrained admissible sets are exploited in definining connectivity and possibility of safe transitions between forced equilibria.  The relative motion planning problem reduces to the graph search for a path between connected equilibria. As in \cite{weiss2014safe}, extensions to the case of multiple control gains appear possible as well as the use of multiple observer gains; details are left to future publications. Connectivity is determined via chance-constrained admissible sets, which allows the consideration of output feedback and Gaussian process and measurement noise while still enforcing obstacle avoidance constraints. The resulting graph was solved using Dijkstra's algorithm, resulting in a fuel-efficient trajectory that satisfied the chance constraints with probability $1-\alpha$.

\bibliographystyle{IEEEtran}
\bibliography{Acc2020}
\end{document}